\documentclass[aps,prl,showpacs,preprint,preprintnumbers]{revtex4-1}
\usepackage{graphicx}
\usepackage[english]{babel}
\usepackage{amsmath,amssymb} 
\usepackage{stmaryrd,txfonts}
\usepackage[sort&compress]{natbib}
\usepackage[dvipsnames]{xcolor}
\usepackage{epsfig}
\usepackage{pslatex}

\usepackage{ifpdf}
\ifpdf
  \usepackage[dvipdfm,colorlinks,hyperindex]{hyperref}
\else
  \usepackage[hypertex,colorlinks,hyperindex]{hyperref}
\fi

  \hypersetup{
    pdftitle = {Optimal Channel Efficiency in a Sensory Network},
    pdfauthor = {Thiago S. Mosqueiro and Leonardo P. Maia},
    pdfkeywords = {neurodynamics, optimal dynamic range, avalanches, entropy},
    colorlinks = true, 
    linkcolor = blue,
    anchorcolor = blue,
    citecolor = blue, 
    urlcolor = blue
}

\begin{document}

\title{Optimal Channel Efficiency in a Sensory Network}

\author{Thiago S. Mosqueiro}
\email{thiago.mosqueiro@gmail.com}

\author{Leonardo P. Maia}
\email{lpmaia@ifsc.usp.br}

\affiliation{Instituto de F{\'{\i}}sica de S{\~{a}}o Carlos,
  Universidade de S{\~{a}}o Paulo, 13560-970 S{\~{a}}o Carlos, SP,
  Brasil}

\date{\today}

\begin{abstract}
We show that the entropy of the distribution of avalanche lifetimes in the Kinouchi-Copelli model always achieves a maximum jointly with the dynamic range. This is noteworthy and nontrivial because while the dynamic range is an equilibrium average measure of the sensibility of a sensory system to a stimulus, the entropy of relaxation times is a purely dynamical quantity, independent of the stimulus rate, that can be interpreted as the efficiency of the network seen as a communication channel. The newly found optimization occurs for all topologies we tested, even when the distribution of avalanche lifetimes itself is not a power law and when the entropy of the size distribution of avalanches is not concomitantly maximized, strongly suggesting that dynamical rules allowing a proper temporal matching of the states of the interacting neurons is the key for achieving good performance in information processing, rather than increasing the number of available units.
\end{abstract}



\maketitle

In the end of the last century appeared the first claims on the \textit{criticality hypothesis}, stating that some biological systems could evolve towards the \textit{edge of chaos} \cite{packard_adaptation_1988,langton_computation_1990,kauffman_coevolution_1991,bak_how_1999,beggs_criticality_2008,mora_are_2011}, the critical surface separating two phases in an abstract space of parameters. The heuristic justification for this hypothesis is that adaptation of biological systems would be guided by selective pressures favoring the optimization of some key attributes, e.g. the capacity of sensing environments. No matter how appealing was the proposal, earlier work \cite{packard_adaptation_1988,langton_computation_1990,kauffman_coevolution_1991} relied only on simulations and the lack of experimental validation combined with some criticisms \cite{mitchell_revisiting_1993} turned down the theory for some time. However, captivating researchers on brain dynamics, the idea acquired a wholly new motivation \cite{chialvo_emergent_2010}. Indeed not only new theoretical evidence of increased computational performance  at criticality showed up recently \cite{bertschinger_real-time_2004,haldeman_critical_2005,kinouchi_optimal_2006, tanaka_recurrent_2008} but mainly the observation of power-law behavior of neuronal avalanches in cortical networks both \textit{in vitro} \cite{beggs_neuronal_2003,beggs_neuronal_2004} and \textit{in vivo} \cite{petermann_spontaneous_2009,10.1371/journal.pone.0014129,Hahn14072010} constituted stronger-than-ever evidence of the relevance of the edge of chaos to the operation of biological systems. There was some debate regarding the proper characterization of the recorded power laws \cite{touboul_can_2010,klaus_statistical_2011} and whether criticality would be the sole explanation for the observed scale invariance \cite{benayoun_avalanches_2010} but this time the criticality hypothesis is standing up criticisms \cite{beggs_criticality_2008,mora_are_2011}. Remarkably, some experiments have explicitly revealed maximized quantities as information transmission \cite{beggs_neuronal_2003,shew_information_2011}, information capacity \cite{shew_information_2011}, synchronizability \cite{yang_maximal_2012} and dynamic range \cite{shew_neuronal_2009} in cortical networks at a critical condition.

The dynamic range is a sensibility measure associated to the high-slope region of a tuning curve (also response function, the stimuli-response relationship characterizing a sensory network), where nearby stimuli can be most easily discriminated since small changes in stimulus lead to high changes in the firing response. Theoretical work \cite{kinouchi_optimal_2006} indicated that the dynamic range should be maximized in sensory systems when the topology of the network was set in a specific condition, critical for the signal propagation among interacting excitable neurons. A beautiful marriage of theoretical prediction with experimental confirmation happened when an optimal dynamic range was found \cite{shew_neuronal_2009} in cortex slice cultures with a proper balance between excitatory and inhibitory interactions achieved through pharmacological manipulation.

In this study, we employ simulations to show that, when the Kinouchi-Copelli (KC) model \cite{kinouchi_optimal_2006} is tuned at the edge of chaos, the Shannon entropy of its avalanche lifetime statistics (hereafter \textit{information efficiency}) is always jointly maximized with the dynamic range, both in the original \cite{kinouchi_optimal_2006} random graph topology and alternatives. We note that the information capacity (the entropy of avalanche size distribution) does not always exhibit such a \textit{critical optimization}. Indeed, we believe that information efficiency rules the behavior of the dynamic range and outweighs by far the relevance of information capacity in determining the information processing properties of a sensory network. Two previous works discussed critical optimization of entropies of avalanches sizes. The first one \cite{ramo_measures_2007} was a theoretical study of measures of information propagation in Boolean networks and the second one was the discovery of optimal information capacity in critical cortical networks \cite{shew_information_2011}. None of them have raised the possibility that the information efficiency could be maximized instead of capacity.

{\em The model.}--- We start giving a concise description of the KC model \cite{kinouchi_optimal_2006}: each of the $N$ neurons is a cellular automaton that can be at states 0 (quiescent/excitable), 1 (excited) or $2,\cdots,m-1$ (refractory states). The neurons are arranged as a weighted undirected graph with mean degree $K$. The sequential transitions $1 \to 2$, $2 \to 3$, ..., $m-2 \to m-1$ and $m-1 \to 0$ are deterministic. On the other hand, the transition $0 \to 1$ happens for neuron $k$ if either (i) a spark of any of its excited neighbors $j$ reaches it, with probability $p_{kj}$ drawn from an uniform distribution in $[0,\mathrm{p_{max}}]$, or (ii) if $k$ gets an external stimulus, modeled by a Poisson process with rate $r$, resulting in an excitation probability $\lambda = 1 - \exp \left( r \Delta t \right)$ at each time interval $\Delta t$. All other transitions are forbidden.

By setting $\mathrm{p_{max}} = 2\sigma/K$, the mean number of excitations an excited neuron could generate in one time step if all its neighbors were quiescent is $\sigma$, namely, the average branching ratio. Given the fraction of excited nodes $\rho_t$ at time $t$, the proper psychophysical response of the system is the average activity $F = T^{-1} \sum_{t=1}^T \rho_t$. Consequently, the response is a function $F=F(r)$ of the stimulus rate $r$. The dynamic range $\Delta$ is defined in decibels as $\Delta = 10 \log (r_{0.9}/r_{0.1})$, where $F(r_x)=F_{\mathrm{min}} + x [ F_{\mathrm{max}} - F_{\mathrm{min}}]$, $x \in [0,1]$, $F_{\mathrm{max}}=F(\infty)$ is the satured response and $F_{\mathrm{min}}=F(r\to0)$ is the spontaneous activity. Kinouchi and Copelli showed that self-sustained activity is possible if $\sigma$ is greater than $\sigma_c=1$ \cite{kinouchi_optimal_2006}, so that $F_{\mathrm{min}}$ plays the role of the order parameter in a phase transition in the neural activity with $\sigma$ as a control parameter. They have also found a critical optimization for $\Delta$.

In \cite{kinouchi_optimal_2006} the authors studied only the Erd\"os-R\'enyi topology (ERT) with a fixed number $NK/2$ of connections and focused on characterizing the maximization of the dynamic range, as did further works on alternative topologies \cite{copelli_excitable_2007,wu_excitable_2007,chavez_dynamics_2011}. They did not dwell on exploring the bursts of activity (avalanches) generated by their model, although stating that critical networks exhibit both large variance of avalanche lifetimes and a power-law distribution for avalanche sizes with the classical exponent $-3/2$ \cite{harris_theory_2002,zapperi_self-organized_1995}.

In this work, we study the avalanches exhibited by the KC model implemented on both the ERT and the Barab\'asi-Albert \cite{barabasi_emergence_1999} topology (BAT). Unless explicitly stated otherwise, the simulations were performed with $N = 10^5$ and $K = 10$. Given a randomly generated representant of a topology, with chosen average connectivity $K$ and average branching ratio $\sigma$, we randomly choose a neuron of the network to be initially excited while all others are quiescent and record both the number $s$ of neurons that get excited due to that single spark and the number $t$ of consecutive generations the network remained active. We repeat this procedure a large number of times in order to get the distributions $\{p_s\}$ and $\{p_t\}$ of the size and the lifetime of an avalanche, respectively. In this setting, there is no role at all for a stimulus rate.

\begin{figure}
\centering
\includegraphics[scale = 0.12]{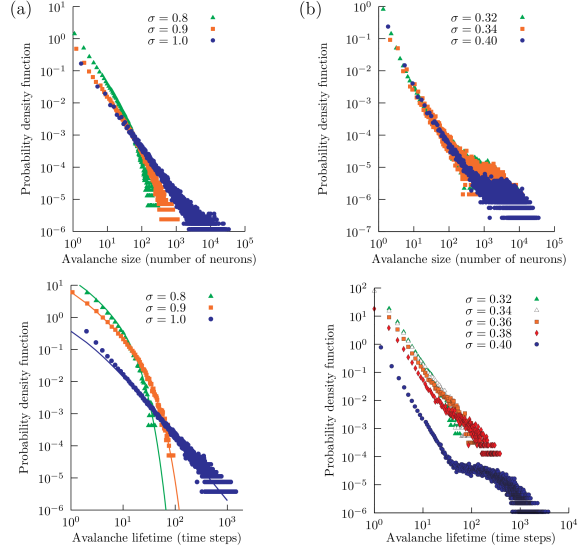}
\caption{Distributions of (top) burst size and (bottom) activity lifetimes for (a) Erd\"os-R\'enyi and (b) Barab\'{a}si-Albert topologies. In (a), power laws emerge for $\sigma_c = 1$ and the solid lines in the bottom result from the theory of branching process (see text). In (b), especially in the bottom, huge avalanches become more frequent when $\sigma = 0.4$. We have evidence that this is the critical condition, despite the lack of a power law (see text).}
\label{fig:av_distros}
\end{figure}

{\em Size and lifetime distributions.}---  Fig. \ref{fig:av_distros} (a) illustrates the critical (at $\sigma_c=1$) emergence of power-law scaling in the bursts of activity, $p_s \sim s^{-1.5}$ and $p_t \sim t^{-1.9}$, for the ERT (exponents estimated with standard techniques \cite{newmancp2005,clausetsiamr2009}). Since the ERT allows the propagation of almost independent branches of activity, this behavior is perfectly compatible with the predictions from the theory of branching processes \cite{harris_theory_2002,zapperi_self-organized_1995}. Indeed, we will describe elsewhere how that formalism predicts the solid lines in the bottom of Fig. \ref{fig:av_distros} (a).

In Fig. \ref{fig:av_distros} (b) we exhibit the avalanche distributions in a BAT. It is harder to estimate these distributions and there is much uncertainty in the size distribution, but the bottom of Fig. \ref{fig:av_distros} (b) shows clearly the presence of a bump in the lifetime distribution for $\sigma=0.4$ ($p_s \sim s^{-2}$ and $p_t \sim t^{-2.9}$ in the first decades) and strongly suggests power-law behavior for slightly smaller values of $\sigma$. A na\"ive analysis based on the criterium of a power law as a signature of critical behavior would favor the latter against the former, but we remark that the observation of power laws in the KC model in this topology demands subsampling to a 2-10\% level (not shown, but see also \cite{10.1371/journal.pone.0014129,PriesemanBMSN}) and below we will present results supporting the ``bumpy'' curve as the critical one. We start discussing the relation among the dynamic range and the entropies of the distributions just described.

\begin{figure}
\centering
\includegraphics[scale=.12]{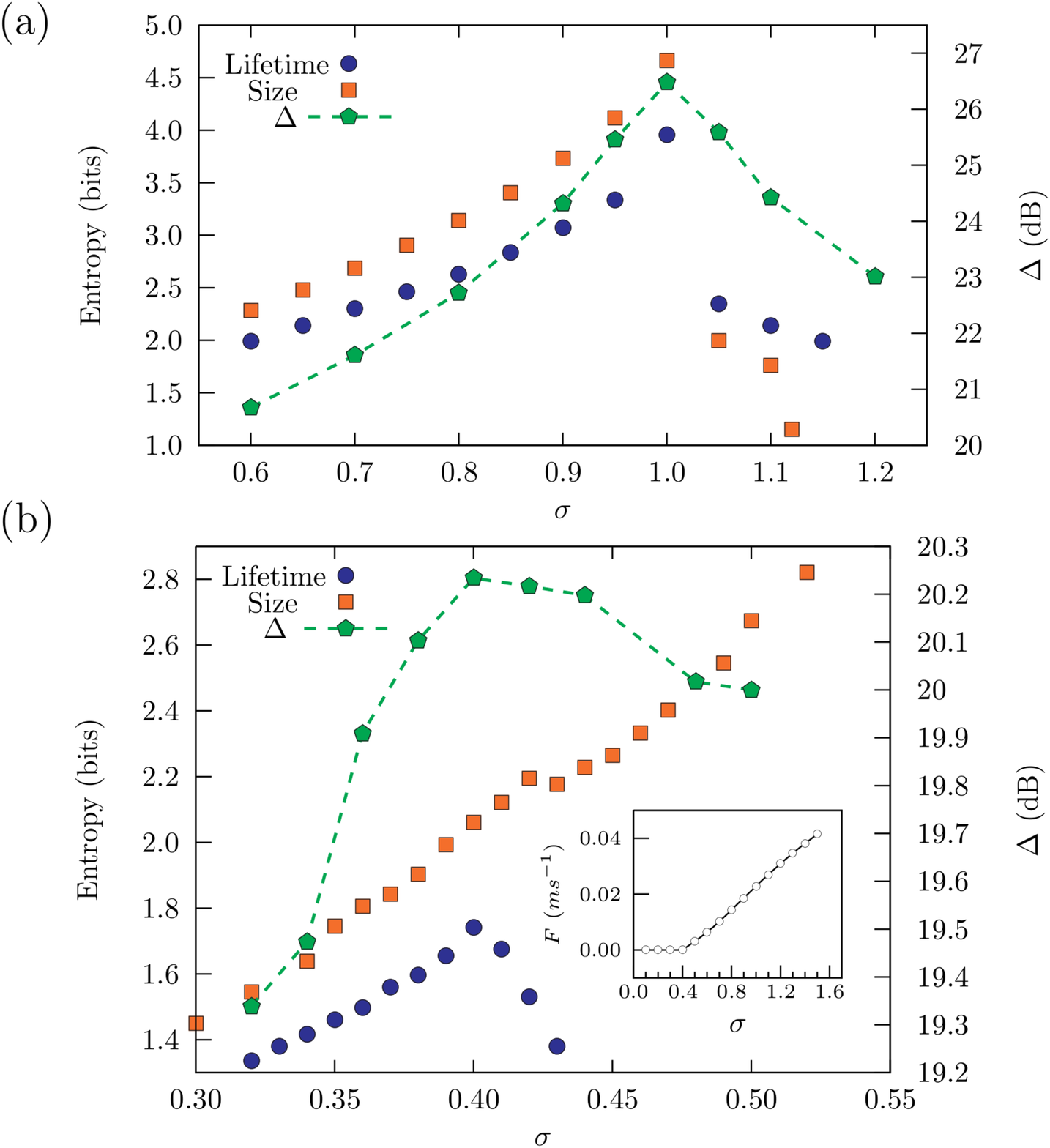}
\caption{Dynamic range ($\Delta$) and entropies of avalanche size and lifetime distributions as functions of the average branching ratio $\sigma$ in (a) Erd\"os-R\'enyi and (b) Barab\'{a}si-Albert topologies. Concomitant critical optimization emerges clearly in (a), but only the lifetime entropy gets optimized jointly with $\Delta$ in (b). The green dashed lines are guides for the eyes. Inset: $F_{\mathrm{min}}$ \textit{vs.} $\sigma$, clearly indicating the critical value $\sigma_c \approx 0.4$.}
\label{fig:entropies}
\end{figure}

{\em Information efficiency.}---  The Shannon entropy $H$ of a distribution $\{p_n\}$,
$$
H (\{p_n\}) = - \sum_{n} p_n \log p_n,
$$
is a standard measure of the uncertainty of a stochastic observable and so it is quite obvious it should be applied to the analysis of avalanches. We will set $n$ as either $s$ or $t$ to indicate which distribution we are talking about. In Fig. \ref{fig:entropies}, we jointly illustrate the behavior of the dynamic range $\Delta$ and of the entropies as functions of $\sigma$, given $K$ and a topology.

It was not surprising to observe in Fig. \ref{fig:entropies} (a) $H (\{p_s\})$ getting maximized jointly with $\Delta$ in the ERT, since two previous studies \cite{ramo_measures_2007,shew_information_2011}, in different contexts, reported such an effect for the information capacity. Moreover, it seems natural to associate the dynamic range to that entropies (as measures of the flexibility of the system in dealing with signals), so that heuristically $H (\{p_t\})$ should exhibit critical optimization following $\Delta$.

Thus, since the critical optimization of $\Delta$ in BAT's has already been stablished \cite{copelli_excitable_2007,wu_excitable_2007}, the lack of a corresponding peak for $H (\{p_s\})$ in Fig. \ref{fig:entropies} (b) came as a complete surprise. Later we will discuss how that happens while $H (\{p_t\})$ and $\Delta$ keep getting optimized jointly, this time at $\sigma=0.4$.

Is this optimization really \textit{critical}? As did the authors of \cite{copelli_excitable_2007}, we claim so invoking as first evidence the plot of the order parameter $F_{\mathrm{min}}$ of the KC model against the control parameter $\sigma$ in the inset of Fig. \ref{fig:entropies} (b). However, our results suggest $\sigma_c=0.4$, while $\sigma_c \approx 0.5$ was estimated in \cite{copelli_excitable_2007}. Despite the much smaller networks simulated in that work, such a divergence demands an even more stringent analysis regarding the critical nature of the position of the peaks in this topology, namely, a tentative of data collapse into a scaling function. We emphasize we are \textit{not} going to take the abscissa of the peak as the location of a critical point \textit{a priori}, since there is no support for such a procedure. Quite the opposite, in \cite{ramo_measures_2007}, for instance, stochasticity makes $H(\{p_s\})$ exhibit a peak away from criticality. 

We also remark that (i) the deviation of $\sigma_c$ from 1 in alternative topologies has been recently explained in terms of a spectral analysis \cite{larremore_predicting_2011, larremore_effects_2011} and nowadays is not surprising at all, (ii) there is no critical optimization for BAT's grown node by node (\textit{i.e.}, the information efficiency and the dynamic range keep behaving jointly even in such a ``charmless'' scenery) and (iii) whatever be the critical point, it is very clear from Fig. \ref{fig:entropies} (b), mainly from the plot of the information efficiency $H (\{p_t\})$, that the joint optimization of $H (\{p_t\})$ and $\Delta$ does \textit{not} happen at the values of $\sigma$ resembling power-law behavior in Fig. \ref{fig:av_distros} (b).


\begin{figure}
\includegraphics[scale=.15]{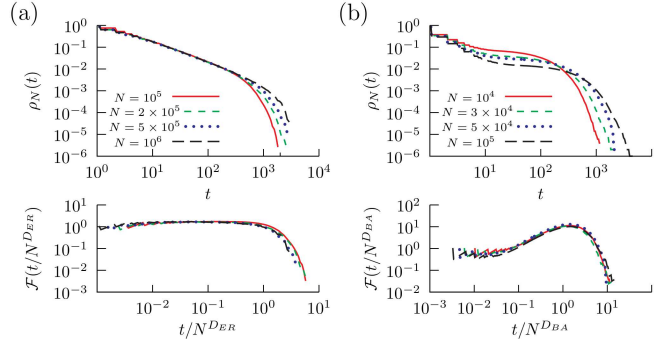}
\caption{Instances of (top) CDF's for several system sizes and (bottom) their collapse into a scaling function for both (a) Erd\"os-R\'enyi and (b) Barab\'asi-Albert topologies. To collapse the CDF's at the proposed critical conditions, we have fit $\gamma_{ER} = 0.9$, $D_{ER} = 0.5$, $\gamma_{BA} = 1.25$ and $D_{BA} = 1$ (see text). The distributions in (b) closely resemble the ones presented recently by Dehghani \textit{et al} \cite{destexhearxiv}.}
\label{fig:collapse}
\end{figure}

{\em Critical optimization.}---  Despite the inset in Fig. \ref{fig:entropies} (b) being a classical signature of a phase transition, the aforementioned displacement of the critical point prompted us to perform a scaling analysis in order to confidently assess the issue of criticality. Let $\rho_N (t)$ be the complementary cumulative distribution function (CDF) of the lifetime of an avalanche when there are $N$ neurons in the KC model (\textit{i.e.}, $\rho_N (t)$ is the probability of the duration of an avalanche surviving for at least $t$ given a network with $N$ units). Given a topology and a value of $\sigma$ thought to be a critical point, we have looked for exponents $\gamma$ and $D$ able to make all the transformed distributions $x^{\gamma} \rho_N (x)$ collapse into a single universal curve $\mathcal{F}$ when plotted against $x/N^D$, regardless the value of $N$. Fig. \ref{fig:collapse} illustrates the success of such endeavour, considering the distributions obtained with $\sigma_c=1$ in the ERT and $\sigma_c=0.4$ in the BAT.

There are two more observations supporting our claim of criticality in the absence of pure power laws in the avalanches (data not shown). Both preliminary results on the collapse of avalanche shapes, as advocated by \cite{sethnanature2001} and implemented in \cite{PhysRevLett.108.208102}, and an analysis of the extinction probability (a branching-process-like study to be discussed elsewhere) point to $\sigma_c=0.4$ in the BAT.

We also report complementary observations regarding data not shown in this paper. In contrast with other contexts \cite{bertschinger_real-time_2004,20053290}, our results are qualitatively robust to changes in the distribution of edge weights (even for constant weigths). Most alternative topologies \cite{dorogovtsev_structure_2000,watts_collective_1998} leading to more realistic degree distributions \cite{eguiluz_scale-free_2005} are qualitatively equivalent to the BAT and exhibit even more pronounced bumps. In the uncorrelated version \cite{catanzaro_generation_2005} of the configuration model \cite{molloy_critical_1995}, bumps are still present in the avalanche distributions but both information capacity and efficiency exhibit critical optimization. This behavior fits in the phenomenology we describe here, but also suggests that the existence of degree correlations in the BAT may be the reason why $H (\{p_s\})$ does not maximize in that topology.

{\em Discussion.}---  Neuroscientists regularly employ the mutual information as a measure of the statistical dependence between stimulus and response, taking into account both specificity (change in stimulus implying change in response) and fidelity (low variability given a stimulus) of responses. Notwithstanding the relevance of such studies, indispensable for the comprehension of neural coding, we remark our aim in this paper is the study of intrinsic, stimulus-free, behavior of sensory systems rather than of patterns of response variability.

The KC model \cite{kinouchi_optimal_2006} revealed that the stimulus-dependant measure $\Delta$ is optimized precisely at $\sigma_c$, a critical point of the self-sustained activity $F_{\mathrm{min}}$, that does not depend at all on the nature of the stimulus. If not a theoretical artifact, that phenomenology suggests that generally the behavior of actual excitable networks (\textit{e.g.}, sensory systems) could be strongly determined by its stimulus-free properties. Accordingly, an optimal dynamic range was observed experimentally at a condition determined in the absence of stimuli \cite{shew_neuronal_2009} and there is further robust evidence that stimulus-evoked activity is strongly dependent on the spontaneous ($r=0$) firing patterns in the cortex (see references in \cite{shew_neuronal_2009}). Therefore, despite it may seem quite plausible \textit{per se} that the sensibility measured by the dynamic range gets optimal when signals do not either fade out fastly (subcritical) or frenetically superimpose themselves (supercritical), there is enough motivation for scrutiny on the microscopic mechanisms leading to improvements in information processing capabilities even in stimulus-free conditions.

We firmly believe to have discovered one such mechanism. It seems fairly intuitive to expect that a sensory system be more flexible if it disposes of greater variability of the duration of the bursts of neural activity it must process. Indeed the support of the avalanche lifetime distribution (the interval of lifetimes with positive probabilities) achieves a maximum jointly with the dynamic range, while noncritical distributions hold for no longer than two decades (see Fig. \ref{fig:av_distros}). We have chosen the information efficiency as the proper measure of that effect because it takes into account not only the magnitude of the reportoire of lifetimes but also a balanced utilization \footnote{As an entropy, the information efficiency gets higher the more spread is the avalanche lifetime distribution on its support. Also on this issue, we remark that the breakdown of power-law scaling in the critical condition occurs by increased concentration of probability mass in the tails of the distributions instead of throwing away large and durable excursions, see Fig. \ref{fig:av_distros}.} of that resource for information transmission (from the olfactory bulb to the cortex, for instance).

It is not evident why such arguments are always valid for the information efficiency but not for the capacity. We speculate that an explanation must rely on the relationship between the microstructure of the network (motifs) and the dynamics of the excitable units. Progress beyond the initial studies on the synchronizability properties of the KC model \cite{chavez_dynamics_2011,rozenblit_collective_2011,yang_maximal_2012} will be probably achieved by deciphering such relationship. As a rule of thumb, greater values of the clustering coefficient \cite{watts_collective_1998} should lead to stronger deviations from the $-3/2$ law. It is also worth mentioning that it has already been suggested \cite{teramae_local_2007} that ``the lifetime distributions of neuronal avalanches may carry rich information about the local cortical circuit structure'' and may exhibit consistent deviations from power-law scaling, while the size distribution would be much more well-behaved. Anyway, despite our praise of the intrinsic activity, it is definitively worth studying stimuli-dependant features of excitable networks by information-theoretic tools like mutual information and transfer entropy.

The notion of criticality without power laws may have wide implications in the interpretation of observations of neuronal avalanches. Recent experiments exhibiting critical optimization \cite{shew_neuronal_2009,shew_information_2011,yang_maximal_2012} have described the phase transitions in terms of a control parameter $\kappa$ resembling $\sigma$ but based on the tacit assumption that neuronal avalanches are pure power laws. Further investigations are necessary to reveal eventual consequences of the breakdown of that hypothesis. Likewise, Ref. \cite{destexhearxiv} employed robust statistical techniques to analyze neuronal avalanches \textit{in vivo} and stand up against critical dynamics. However, the CDF's they present are very similar to Fig. \ref{fig:collapse} (b), so that probably their data is ruling out power-law scaling, but not criticality. Finally, distributions pretty much like the ones in Fig. \ref{fig:av_distros} (b) have been recently observed in high-resolution experiments \textit{in vitro} \cite{PhysRevLett.108.208102} and the bumps were no obstacle for a remarkable data collapse constituting very compelling evidence of critical behavior in brain dynamics.

Summarizing, we studied the avalanches in Kinouchi-Copelli model in a first attempt to figure out detailed mechanisms of information transmission in cortical networks. We discovered that, in a critical point, the entropy of avalanche lifetime statistics (information efficiency) is always maximized jointly with the dynamic range, an important measure of information transmission extracted from the psychophysical tuning curves. Our findings fit in the discussions regarding the role of criticality in information processing \cite{beggs_criticality_2008,mora_are_2011} and the relationship of long bursts of activity with the dynamic range \cite{shew_neuronal_2009}, specially because they suggest critical behavior without pure scale invariance.

{\em Acknowledgements.}  We acknowledge the contribution of an anonymous referee that brought reference \cite{PhysRevLett.108.208102} to our attention and gave suggestions that significantly improved the paper. Thiago S. Mosqueiro acknowledges CAPES for financial support. The work of Leonardo P. Maia was supported by FAPESP grant No. 2010/20446-5.



%

\end{document}